\documentstyle[psfig,floats,preprint,aps,prl,array]{revtex}

\ifx\undefined\psfig\def\psfig#1{ }\else\fi

\def\ofog1{\hline}
\def\be{\begin{equation}}
\def\ee{\end{equation}}
\def\bea{\begin{eqnarray}}
\def\eea{\end{eqnarray}}
\def\ba{\begin{array}}
\def\ea{\end{array}}

\def\E{{\rm E}}
\def\T{{\rm T}}
\def\a{\alpha}
\def\b{\beta}
\def\V{{\rm V}}
\def\f{\frac}
\begin{document}
\ifpreprintsty\else
\twocolumn[\hsize\textwidth%
\columnwidth\hsize\csname@twocolumnfalse\endcsname
\fi

\preprint{IPM-98-17}
\title{ Jahn-Teller Effect in Diamond-like Carbon
             }
\author{
{\bf M.A. Vesaghi}${}^{a}$\footnote
{e-mail address: vesaghi@physic.sharif.ac.ir}
and {\bf A. Shafiekhani}${}^b$\footnote{e-mail address:
ashafie@theory.ipm.ac.ir} 
}
\vspace{0.5 cm}
\address{        ${^a}$ Dept. of Physics,
               Sharif University of Technology,\\
               P.O.Box: 9161,Tehran 11365, Iran\\
${^b}$ 
Institute for Studies in Theoretical Physics and Mathematics\\
            P.O.Box: 5531, Tehran 19395, Iran
}
\maketitle
\begin{abstract}
\leftskip 2cm
\rightskip 2cm
The Jahn-Teller effect used semi-theoretically to analyse
 UV-visible spectra of 
the diamond like-carbon films on Si substrates. By deconvolution of 
UV-visible absorption spectra of the typical films, different absorption 
lines found. For each sample, six lines which were related to 
$\V^\circ$ by Jahn-Teller effect are distinguished. 
Our theoretical approach based on distortion of the $\V^\circ$ as the 
consequence of the stress in the films, is in agreement with  
experimental results. The shift of the Jahn-Teller lines 
toward the zero phonon line of $\V^\circ$ were found in accordance to stress 
decrease in the films.
The difference between room temperature and low
temperature spectra is discussed.
We found that the splitting of the excited state of $\V^\circ$ in the films 
under stress is twice as large as that of the ground state.
\end{abstract}


\ifpreprintsty\else\vskip1pc]\fi
\narrowtext

\section{Introduction}
There is a growing interest in vacancies and vacancy related effects in DLC,
a-C and a-C:H films from theoretical and experimental points of view. 
Despite large improvement in the preparation techniques for 
these type of films, a 
large number of vacancies is present in such films \cite{vdlc}. 
This is due to the 
nature of preparation since practically it is not possible to make films free 
of vacancies.

Different varieties of vacancies exist in these films but the main 
ones are neutral-single vacancy ${\rm V}^\circ$, one carbon with its four  
valance electrons is missing, and negatively charged-single 
vacancy ${\rm V}^-$, one carbon with its three valance electrons is 
missing. Such vacancies exist in natural diamond or are
made in it by neutron, electron or gamma bombardment \cite{vdd,cw,gpn}.
These vacancies have peculiar absorption, photoluminescence (PL) and 
cathodluminescence (CL) emission .
All the PL, CL and absorption spectra show many feature, including
different peaks and shoulders \cite{pl,cl,rcf,vs}. The GR1 band with zero 
phonon line at 1.673 eV and ND1 with zero phonon line at 
3.149 eV are associated with $\V^\circ$ \cite{ck} 
and $\V^-$  \cite{lo}, respectivly.

Much work is done on the effect of uniaxial stress in natural diamond 
to explain lines of GR1 band \cite{ck,dp,dav,j.t}. It is reported 
that for films there is a strong correlation between PL intensity 
and stress \cite{fs} and also, between the line-width of zero phonon 
line and film strain \cite{cks,rcf}

In recent years, considerable attention has been paid to the degenerate
electron-lattice interaction in molecules and defect centres in solids,  
called Jahn-Teller effect (JTE) \cite{dm,del,lo}. 

In this paper JTE is used phenomalogically to explain
 our experimental results on absorption spectra of DLC film on
 silicon substrate. The nature of stress in films and its distortion
effect is explained. The electronic states splitting of $\V^\circ$ under 
stress are obtained.
The relative intensity change of room temperature spectra to that of low
temperature spectra also is explained.
\section{Experiment}

DLC films were made from liquid gas (60\% Butane and 40\% Propane) on
$10\times 20\times 1 {\rm mm}$ silicon substrates.
The growth conditions were:\\
Gas pressure: $85\times 10^{-3}$ Torr\\
Substrate temperature:  $200{}^\circ$ C\\
Direct voltage:  450 V\\
Deposition time:  different for different samples.\\
Different deposition time means different thickness of films.
The absorption of the films were measured by double beams 
spectrometer at room temperature and in the reflection mode.

The absorption spectra of typical films in UV-visible region 
(200-1200 nm) are shown in Fig. ~\ref{fsi}. As it is evident from 
this figure the absorption peaks are at 
different wavelength for different samples. They have moved in accordance
to the deposition time of growth, showing that the peak displacement
depends on thickness.
\begin{figure}\label{fsi}
\centerline{\psfig{figure=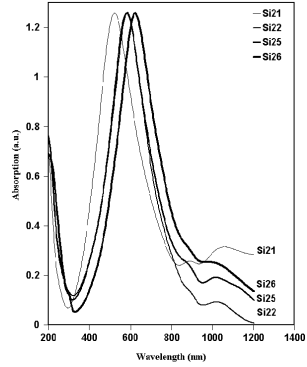,width=.24\columnwidth}}
\caption{Room temperature UV-visible spectra of films deposited
at 85 mtorr and 200${}^\circ$C for samples (Si21) 1 hour
, (Si22) 2.5 hour,(Si25) 2 hour, and (Si26) 4 hour on Si-substrates.
}
\end{figure}
The spectra were deconvoluted by normalized Gaussian with accuracy 
better than 99.95\%. The resultant deconvolution for mentioned samples are 
shown in Fig. ~\ref{dc}.

Some of these lines which were found by deconvolution, had been reported 
before and some had not \cite{vdlc,vdd,cw,pl,cl,vs,fs,dp,dav,cks,spect}. 
We will come to this point later.
\widetext
\begin{figure}\label{dc}
\centerline{\psfig{figure=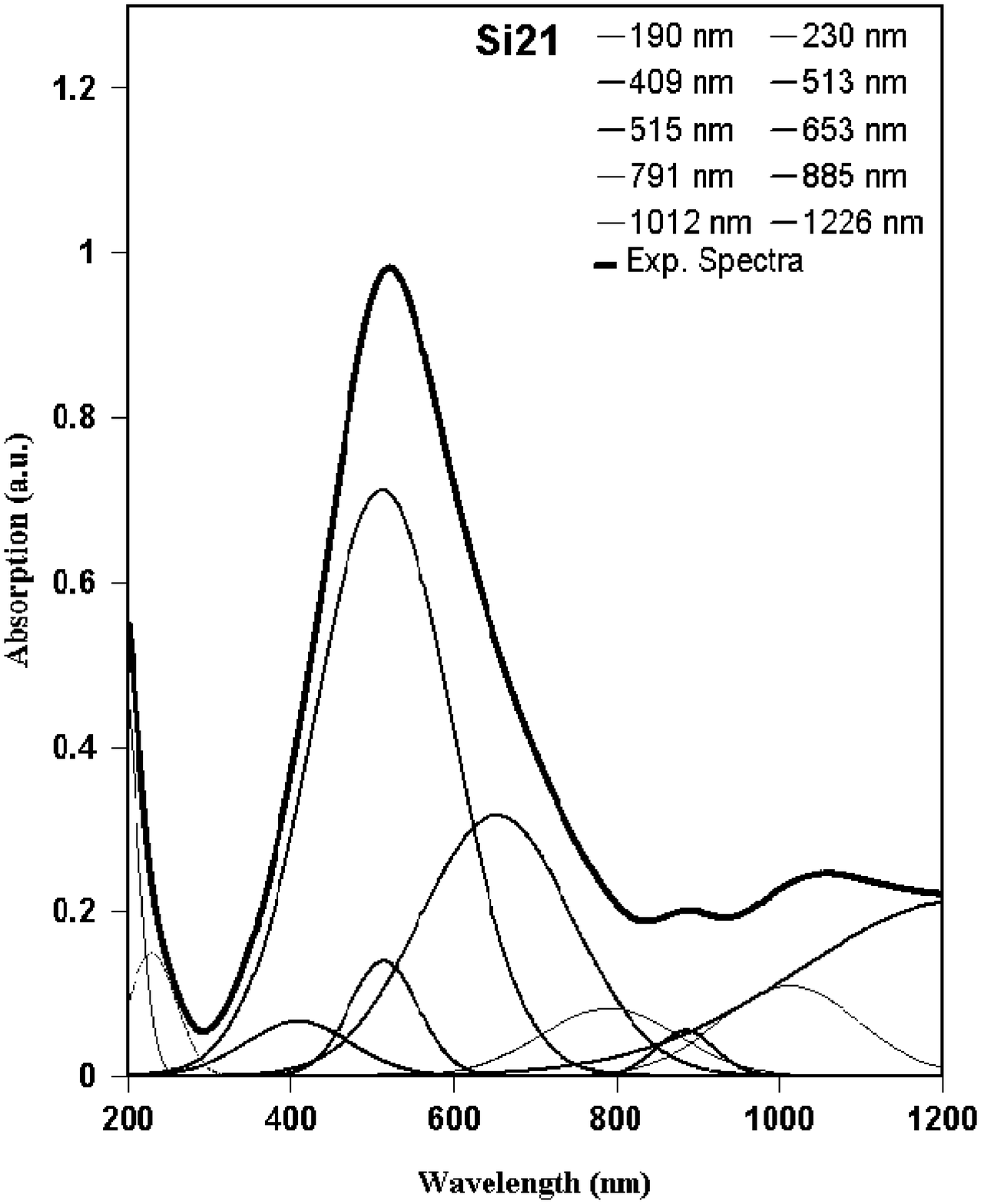,width=.24\columnwidth}
            \psfig{figure=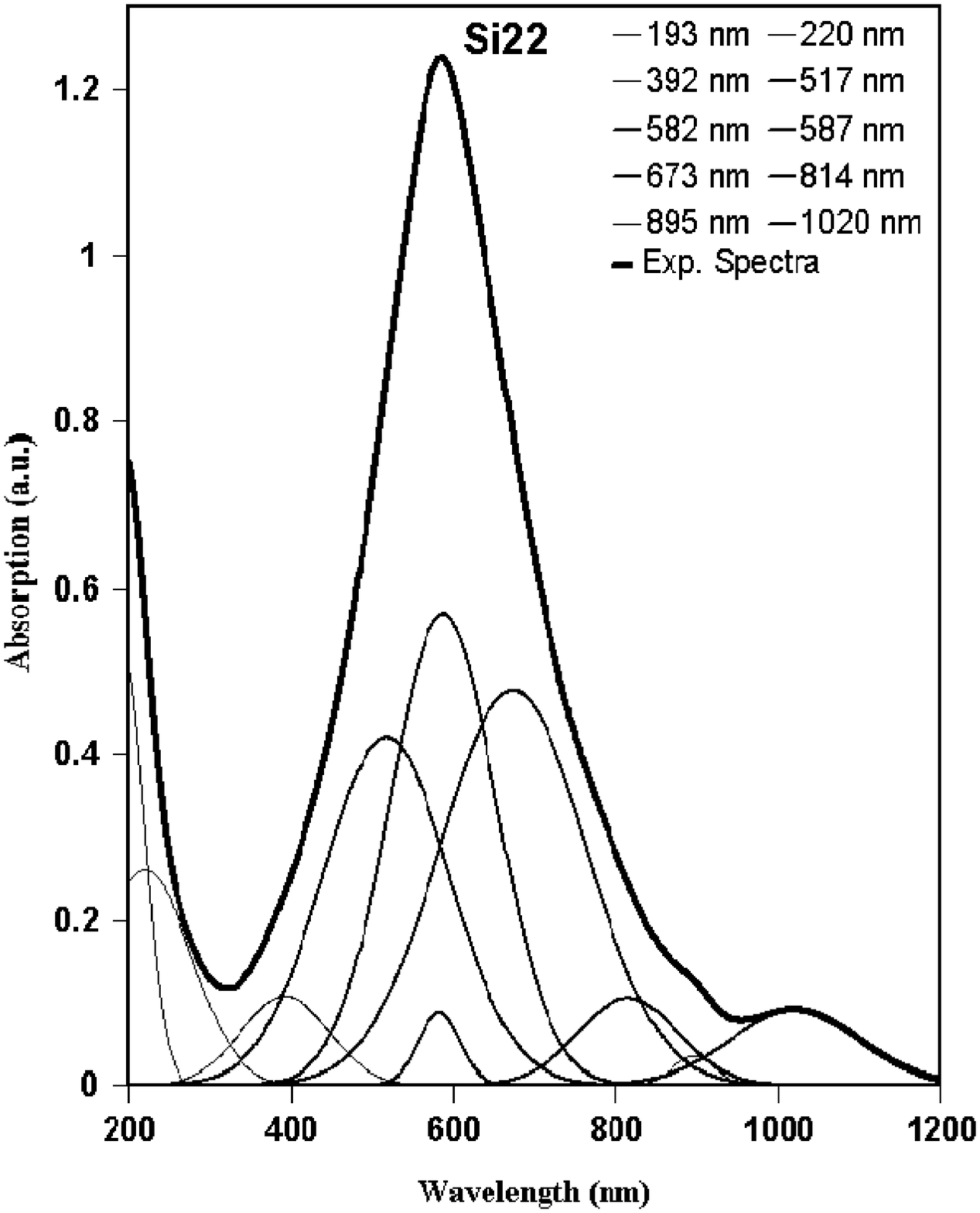,width=.24\columnwidth}
            \psfig{figure=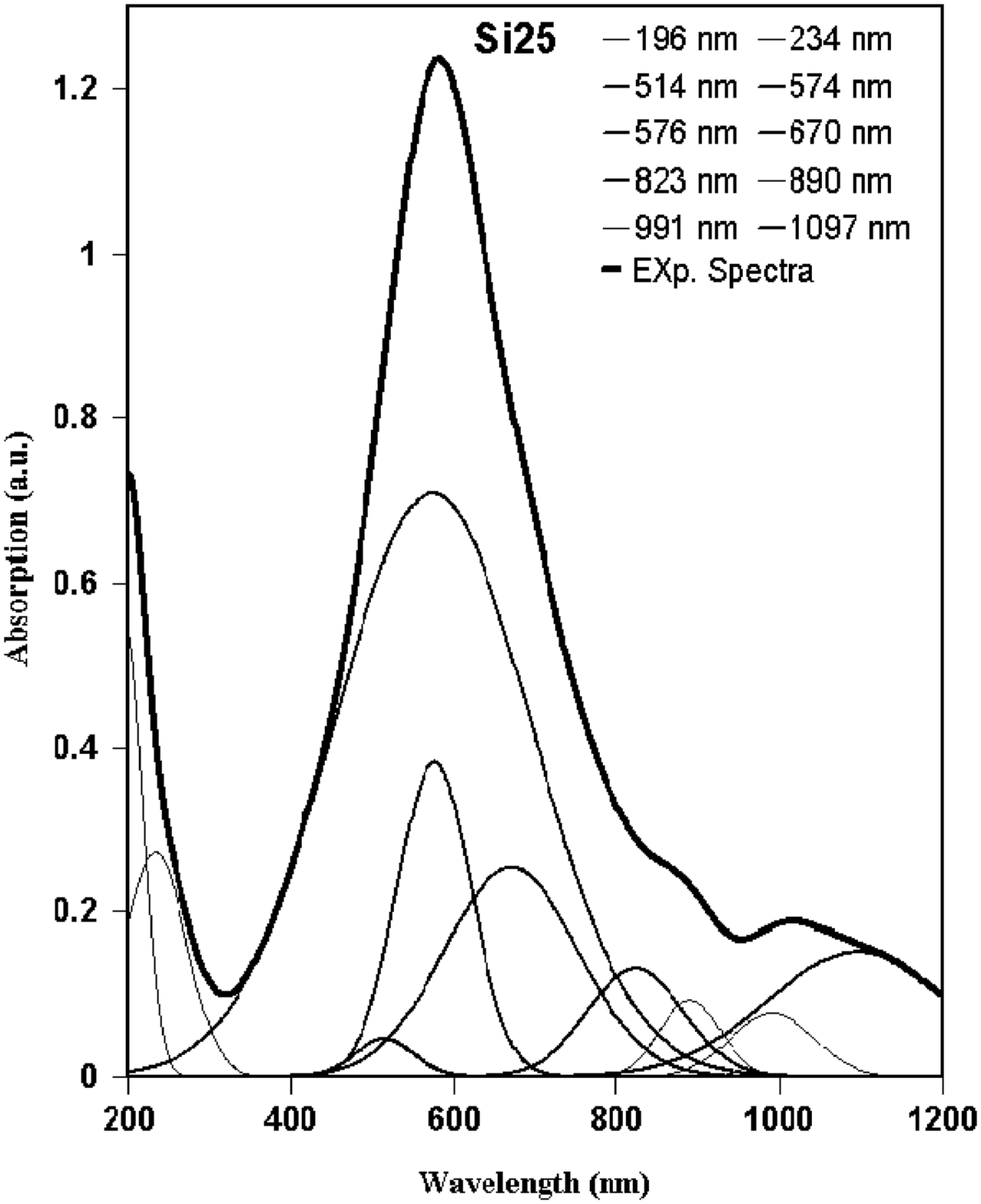,width=.24\columnwidth}
            \psfig{figure=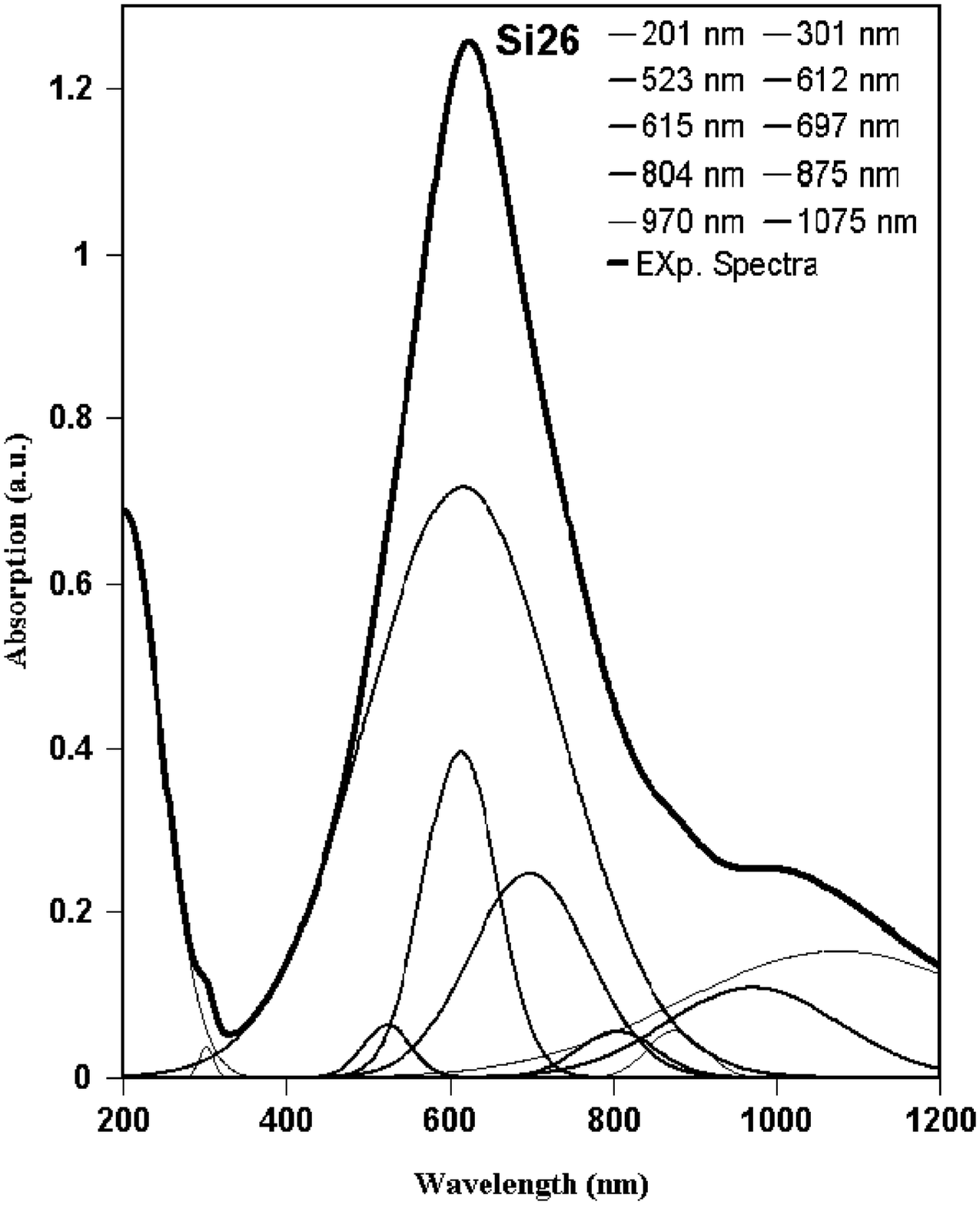,width=.24\columnwidth}}
\caption{Gaussian deconvolution of UV-visible spectra of "Si21", "Si22",
"Si25" and "Si26". For each sample, curve by
{\vbox{\hrule width0.4cm height1pt}}
show the experimental spectra, curves by
{\vbox{\hrule width0.4cm height.2pt}}
show the unrelated lines and curves by
{\vbox{\hrule width0.4cm height0.5pt}}
show the Jahn-Teller lines.
}                               
\end{figure}
\narrowtext
\section{Jahn-Teller effect}

In $\V^\circ$ vacancy of diamond structure there are four ${\rm sp}^3$
electrons rigidly connected to the atoms surrounding the empty space of 
the vacant atom.
Hence, there is a molecule with four atoms, each with one ${\rm sp}^3$ 
electron sitting at the A, B, C, and D corners of the cubic unit cell 
of the diamond structure as shown in Fig. ~\ref{vz} \cite{ck}. 

Degenerate electron state of nonlinear molecule such as $\V^\circ$  and 
$\V^-$  as stated by Jahn and Teller \cite{jt} are not stable and the 
degeneracy of these molecules have to be removed or reduced for stability.
With no distortion, such a molecule has all the symmetry of the diamond 
structure except the translational symmetry. The point symmetry of the
diamond lattice is ${\bf\rm T_d}$. 
The one electron states of this molecule have $a_1$ and $t_2$ symmetry. 
With those ${\rm sp}^3$ shown in Fig. ~\ref{vz}, we have
\be\ba{l}
a_1=\f{1}{2}[\psi_A+\psi_B+\psi_C+\psi_D]\\
t_{2x}=\f{1}{2}[\psi_A-\psi_B+\psi_C-\psi_D]\\
t_{2y}=\f{1}{2}[\psi_A-\psi_B-\psi_C+\psi_D]\\
t_{2z}=\f{1}{2}[\psi_A+\psi_B-\psi_C-\psi_D].
\ea\ee
where $\psi_i$ is ${\rm sp}^3$ wave function at $i$th site.
The degenerate electron states
correspond to the representations E, ${\rm T_1}$ and ${\rm T_2}$
of this group.

\begin{figure}\label{vz}
\centerline{\psfig{figure=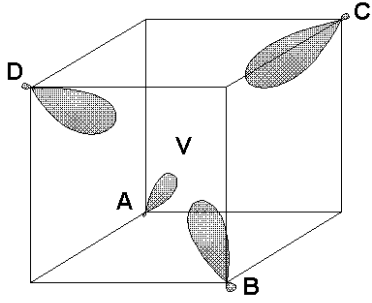,width=.3\columnwidth}}
\caption{$\V^\circ$ vacancy of diamond structure. Each ${\rm sp}^3$ electron are sited in 
A, B, C and D corners.
}
\end{figure}
In this molecule electron-electron correlation leads to a multiple
structure with ${}^1{\rm E}(a_1^2t_2^2)$ being the lowest energy state 
observed for $\V^\circ$. Transition to the state ${}^1{\T_2}(a_1^2t_2^2)$ 
is then associated with GR1 \cite{ck}.

The ground state,  ${}^1{\E}$ 
and the first excited state, ${}^1{\T_2}$  of this molecule have
two and three fold symmetry, respectively. 
Therefore, GR1 bond is sum of six transition lines.  
Under a distortion such as stress, the symmetry of this molecule will be 
broken, hence, the degeneracy of the levels will be
resolved and there will be five splitted levels, two for ground state,  
${}^1{\E}$, and three for exited state, ${}^1{\T_2}$;
\bea
{\rm E}&\rightarrow&  {\rm E}\pm \a,\cr
{\rm T}&\rightarrow&  {\rm T}\pm \b, {\rm T}+ 2\b
\eea
where E and T are the energies of ${}^1{\E}$ and ${}^1{\T_2}$ levels, respectively
\cite{eng}.
With such splitted levels, there are six transitions as shown in Fig. ~\ref{sp}.
\begin{center}
\hspace{3cm}
\unitlength 1.00mm
\linethickness{0.4pt}
\begin{figure}
\begin{picture}(41.00,56.00)(-41,0)
\put(39.00,6.00){\line(-1,0){15.00}}
\put(24.00,6.00){\line(-1,1){5.00}}
\put(19.00,11.00){\line(-1,0){15.00}}
\put(19.00,11.00){\line(1,1){5.00}}
\put(24.00,16.00){\line(1,0){15.00}}
\put(19.00,36.00){\line(-1,0){15.00}}
\put(39.00,26.00){\line(-1,0){15.00}}
\put(24.00,26.00){\line(-1,2){5.00}}
\put(19.00,36.00){\line(1,2){5.00}}
\put(24.00,46.00){\line(1,0){15.00}}
\put(19.00,36.00){\line(1,4){5.00}}
\put(24.00,56.00){\line(1,0){15.00}}
\put(26.00,6.00){\vector(0,1){20.00}}
\put(28.00,6.00){\vector(0,1){40.00}}
\put(30.00,6.00){\vector(0,1){50.00}}
\put(11.00,11.00){\line(0,1){10.00}}
\put(11.00,26.00){\line(0,1){10.00}}
\put(1.00,36.00){\makebox(0,0)[rc]{${}^1{\rm T}$}}
\put(1.00,11.00){\makebox(0,0)[rc]{${}^1{\rm E}$}}
\put(11.00,24.00){\makebox(0,0)[cc]{1.673 ev}}
\put(41.00,56.00){\makebox(0,0)[lc]{${}^1{\rm T}+2\b$}}
\put(41.00,46.00){\makebox(0,0)[lc]{${}^1{\rm T}+\b$}}
\put(41.00,26.00){\makebox(0,0)[lc]{${}^1{\rm T}-\b$}}
\put(41.00,16.00){\makebox(0,0)[lc]{${}^1{\rm E}+\a$}}
\put(41.00,6.00){\makebox(0,0)[lc]{${}^1{\rm E}-\a$}}
\put(11.00,1.00){\makebox(0,0)[cc]{{\bf (a)}}}
\put(32.00,1.00){\makebox(0,0)[cc]{{\bf (b)}}}
\put(33.00,16.00){\vector(0,1){10.00}}
\put(35.00,16.00){\vector(0,1){30.00}}
\put(37.00,16.00){\vector(0,1){40.00}}
\end{picture}
\caption{{$\V^\circ$} transitions: (a)- nondisturbed, (b)- disturbed.} 
\end{figure}
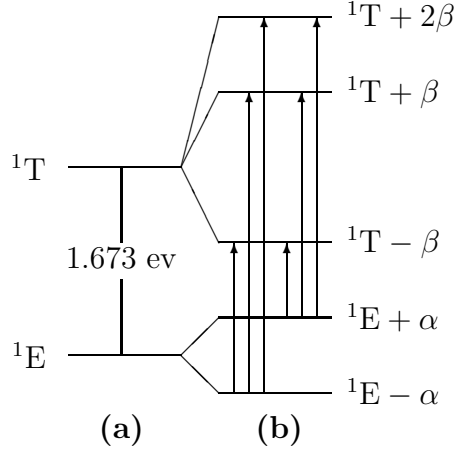
\label{sp}
\end{center}
The energies for this transitions are;
\bea\label{enr}
(\T-\E)+(2\b+\a)&,& (\T-\E)+(2\b-\a),
\cr
(\T-\E)+(\b+\a)&,& (\T-\E)+(\b-\a),
\cr
  (\T-\E)+(-\b+\a)&,& (\T-\E)+(-\b-\a).
\eea
Using the experimental information of any two lines, one is able to calculate 
$\a$ and $\b$. Then using $\a$ and $\b$ so obtained, the transition energy of four other 
lines could be predicted in this approach.
\section{Theoretical results}

As it is apparent from deconvolution result of each sample, there are two
lines very close to each other, namely 513 and 514 nm lines for "Si21", 
583 and 586 nm lines for "Si22", 574 and 576 nm lines for "Si25" and 612 
and 615 nm for "Si26". The only rational possibility for this to happen 
is the case when the transition energy of $\E-\a\rightarrow\T+\b$ is 
very close to the transition energy of $\E+\a\rightarrow\T+2\b$. In this case 
$\b=2\a$. Physically this means that the
splitting of the exited state is twice the splitting of the ground state.

($\T-\E$) is the transition energy of non-disturbed ${\V^\circ}$ 
which is 1.673 eV.
 Using this value and the experimental resultant explained above, 
 $\a$ and $\b$ are calculated for each sample. 
Inserting the obtained values of $\a$ and $\b$ in Eqs. (~\ref{enr}), 
the Jahn-Teller transition lines for each sample were predicted. 
These results are summarized in Table I. 
\narrowtext
\oddsidemargin -5 mm
\evensidemargin -5 mm
\begin{table}
\begin{center}
\caption{Theoretical and experimental values, first row sample no., second row 
is deposition time (D.T.) in hour, third row is theoretical
 energy (T.E.), forth row is theoretical calculated wavelength (T.W.),
 fifth row is experimental wavelength, sixth row is Jahn-Teller lines 
 assigned to GR1 band (JTLGR1) and last row is relative error.
}
\begin{tabular}{llllllllllllllllllllllllllll}
\tableline
\tableline
Sample&&&\multicolumn{11}{c}{Si21}&&&&\multicolumn{11}{c}{Si22}\\
\tableline
D.T. (hour)&&&\multicolumn{11}{c}{1.0}&&&&\multicolumn{11}{c}{2.5}\\
$\a$ (eV)&&&\multicolumn{11}{c}{0.246}&&&&\multicolumn{11}{c}{0.149}\\
T.E. (eV)&&&2.90&&2.41&&2.41&&1.92&&1.43&&0.93&&&&
2.42&&2.12&&2.12&&1.82&&1.52&&1.23\\
T.W. (nm)&&&427&&514&&514&&646&&869&&1327&&&&
513&&585&&585&&681&&814&&1011\\
E.W. (nm)&&&409&&513&&515&&653&&885&&1227&&&&
517&&582&&587&&673&&814&&1020\\
JTLGR1&&&GR1a&&GR1b&&GR1c&&GR1d&&GR1e&&GR1f&&&&
GR1a&&GR1b&&GR1c&&GR1d&&GR1e&&GR1f\\
\% Error&&&4.10&&0.16&&0.16&&1.12&&1.84&&7.57&&&&
0.86&&0.26&&0.26&&0.13&&0.00&&1.00\\
\tableline \\
\end{tabular}
\end{center}

\begin{center}
\begin{tabular}{llllllllllllllllllllllllllll}
\tableline
\tableline
Sample&&&\multicolumn{11}{c}{Si25}&&&&\multicolumn{11}{c}{Si26}\\
\tableline
D.T. (hour)&&&\multicolumn{11}{c}{2.0}&&&&\multicolumn{11}{c}{4.0}\\
$\a$ (eV)&&&\multicolumn{11}{c}{0.161}&&&&\multicolumn{11}{c}{0.116}\\
T.E. (eV)&&&2.48&&2.16&&2.16&&1.83&&1.51&&1.19&&&&
2.25&&2.02&&2.02&&1.79&&1.56&&1.33\\
T.W. (nm)&&&500&&575&&575&&676&&820&&1042&&&&
551&&614&&614&&693&&796&&935\\
E.W. (nm)&&&514&&574&&576&&671&&823&&1097&&&&
523&&613&&615&&697&&804&&970\\
JTLGR1&&&GR1a&&GR1b&&GR1c&&GR1d&&GR1e&&GR1f&&&&
GR1a&&GR1b&&GR1c&&GR1d&&GR1e&&GR1f\\
\% Error&&&2.76&&0.17&&0.17&&0.81&&0.35&&5.23&&&&
5.03&&0.23&&0.23&&0.55&&1.03&&3.77\\
\tableline
\tableline
\end{tabular}
\end{center}
\end{table}
\evensidemargin 0 mm
\oddsidemargin 0 mm 
\section{discussion}

In process of making the film, two different stages can be distinguished.
In the first stage, deposition starts as islands at different 
points on substrate and at different times. At the second stage, 
these islands grow and reach each other and then the film grows as a film. 
At the early stage of the film growing process, there are mis-matchings 
in domain walls of islands and also, interface of film and substrate. 
Such mis-matching produce strong stress in the film. This stress  
causes the symmetry breaking and splitting of the energy levels.

As the film thickness increases, the effect of the bulk will  
prevail and only the stress in the interface will be strong. 
Therefore the resultant stress in the film decreases, and consequently 
the effects related to stress such as energy levels splitting decrease. 
Figures (~\ref{fsi}) and (~\ref{dc}) are evidences for this effect. 
As the deposition time increases, meaning that the thickness increases, 
the JT related lines move 
toward zero-phonon line of ${\rm V}^\circ$, 741 nm, Fig ~\ref{zpl}.
\begin{figure}\label{zpl}
\centerline{\psfig{figure=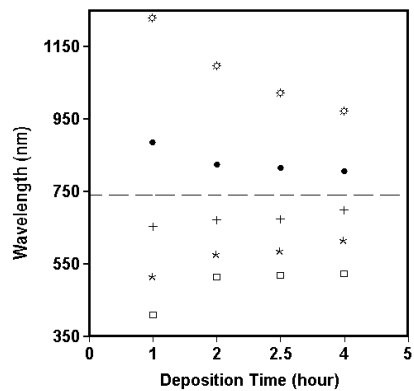,width=.4\columnwidth}}
\vspace{-10 mm}
\caption{Shift of Jahn-Teller lines toward zero phone line of $\V^\circ$, 
 dash line 741 nm, where $\Box$ GR1a, $\star$ GR1b and GR1c, + GR1d, 
 $\bullet$ GR1e and $\ast$ GR1f are.}
\end{figure}
In Table I theoretical and experimental wavelengths of the JT
related lines are compared. The last row shows the good agreement 
between experiment and our theoretical approach.
 
Another important point which deserves attention, is the relative
intensity of the two lines with equal energy. One of these lines 
is the transition from lower level of the ground state, and the other 
is the transition from the higher level of the ground state, which is 
thermally less populated than the first one.

Up to now, in addition to our results, many lines are reported 
by different groups on irradiated and non-irradiated natural 
diamond and carbon films. Similar to ours some of these lines 
are related to vacancies under stress. 
In low-temperature, the transitions from upper ground state are 
weak. In PL and CL also the related transitions to the lower ground 
state are small.
\section{conclusion}

Upon deconvolution of UV-visible spectra many hidden lines appeared which 
 we  considered them for analysis of the spectra. Those non apparent lines 
 are not the less important ones. 

We believe the lines seen at different wavelengths in our results and others 
are due to the vacancy based on JTE. The strength of the effect depends on 
the growth conditions and thickness of the films. 

The results of our phenomenological approach are in good agreement with our 
experimental results. Therefore, this approach could be used to analyse the 
physical properties of carbon films.  

First excited state splitting of $\V^\circ$ in the films under stress 
is two times of ground state splitting in contrast to some theoretical 
models \cite{dm}.

\nopagebreak

\end{document}